\documentclass[aps,prd,showpacs,preprintnumbers,nofootinbib,floatfix,floats,groupedaddress,twocolumn]{revtex4}
\usepackage{bm}
\usepackage{latexsym}
\usepackage{dcolumn}
\usepackage{amsmath,amsfonts,amssymb}
\usepackage{graphicx,epsfig}
\usepackage{color}
\usepackage[active]{srcltx}%DO NOT DELETE
\usepackage{subfig}
\usepackage{slashed}
\usepackage{tikz}
\usetikzlibrary{shapes,snakes}
\def\nn{\nonumber}

\def\l{\left}
\def\r{\right}
\def\DM{\mathrm{d}}

\newcommand{\gae}{\lower 3pt \hbox{$\,\, \buildrel {\scriptstyle >}\over {\scriptstyle
\sim}\,\,$}}
\newcommand{\lae}{\lower 2pt \hbox{$\, \buildrel {\scriptstyle <}\over {\scriptstyle
\sim}\,$}}
\reversemarginpar
\def \lp { \ell_\mathrm{\textsf{UV}} }
\def \rhov {\rho_{0}}
\def \lr { \ell_\mathrm{\textsf{IR}} }

\def \Rcomb {\mathrm{\textsf R}}

\def\eq#1{{Eq.~(\ref{#1})}}

\begin{document}

\title{Vacuum energy in freely falling frames and spacetime curvature
}

 \author{Dawood Kothawala}
 \email{dawood@physics.iitm.ac.in}
 \affiliation{Department of Physics, Indian Institute of Technology Madras, Chennai 600 036}

\date{\today}
%\date{October 10, 2015}
\begin{abstract}
\noindent
The structure of quantum vacuum in presence of gravity, and the corresponding vacuum energy density $\rho_{v}$, is expected to depend on the coupling between the UV scale $\lp$ and spacetime curvature. We determine this coupling in an arbitrary freely falling frame characterised by it's geodesic tangent $\bm U(\tau)$. We show that local vacuum modes within a small causal diamond based on $\bm U(\tau)$, whose size is set by wavelength of the modes,
%with wavelength less that the curvature length scale
%
generically give a contribution $\rhov$ to $\rho_v$ which, to leading order, scales as: $\rhov = \l( \pi \hbar c/2 \r) \Rcomb \, \lp^{-2}$, where the curvature term $\mathrm{\textsf R}=\alpha R_{ab} U^a U^b + \beta R$, and $(\alpha, \beta) \in \mathbb{R}$ are constants. The genericness of this result arises from the fact that, although the modes may reduce to Minkowski plane waves along $\bm U(\tau)$, the stress-energy tensor $T_{ab}$, since it depends on derivatives of the modes, does not reduce to it's Minkowski value on $\bm U(\tau)$. 
We discuss implications of our result for vacuum processes in freely falling frames, particularly in connection with certain aspects of the cosmological constant and horizon entropy. 
%We also define a dimensionless quantity 
%$\mathrm{\textsf Q}(\ell)=(\hbar c)^{-1} \rhov V_{\Diamond}(\ell)$, where $V_{\Diamond}(\ell)$ represents volume of, say, a causal diamond of size $\ell$ based on the geodesic, and observe that $\mathrm{\textsf Q}(\lp)$ and 
%$\mathrm{\textsf Q}(1/\sqrt{|\Rcomb|})$
%can provide a common explanation for: (i) the small value of the cosmological constant, and (ii) the large value of vacuum entropy and it's {\it quadratic} rather than {\it cubic} dependence on size of a spacetime region.
\end{abstract}

\pacs{04.60.-m}
\maketitle
\vskip 0.5 in
\noindent
\maketitle
%%%%%%%%%%%%%%%%%%%%%%%%%%%%%%%%%%%%%%%%%%%%%%%%%%%%%%%%%%%%%%%%%%%%%%%%%%%%%%%%%%%%%%%
\section{Introduction} \label{sec:intro} 
%{\bf Introduction.}
The coupling of quantum vacuum energy with gravity, or equivalently, with the spacetime curvature, is widely considered to be of utmost significance for a proper understanding of several technical and conceptual aspects of semiclassical and even quantum gravity. It is also of particular significance in cosmology \cite{cc-review-jm}. The fact that our universe is expanding at an accelerating rate remains one of the most important deductions to have come out of the observed cosmological data over the past few decades. As has been already noted in the literature \cite{geom-mean-rel}, the magnitude of energy density that can explain this acceleration has a value which is of the order of $1/\lp^2 \lr^2$, where $\lp$ and $\lr$ are UV cut-off length scale and the Hubble length (the IR length scale) respectively. One obvious source of this energy density could be the zero-point energy of quantum vacuum. However, naturalness arguments based naively on the local structure of the quantum vacuum would suggest a value of $\sim 1/\lp^4$ in addition to a (sub-dominant) IR contribution $1/\lr^4$ -- obviously, neither of these can explain the observed value
\footnote{This is one of the aspects of the cosmological constant problem; for recent reviews, see \cite{cc-review-jm}. For earlier reviews, see \cite{cc-reviews}.}.
This has led to the suggestion that one might require a complete framework of quantum gravity to fully explain the observed acceleration of the universe; for example, as a low energy relic of some non-perturbative quantum gravitational effects \cite{sorkin}. And indeed, this seems like the most probable scenario \cite{dk-tp-cheshire, viqar}. However, as exciting and appealing this possibility is by itself, it is equally important to clarify the reasons for the failures of attempts to explain the observed acceleration within the context of our well tested theories of general relativity and quantum field theory. Naively, this failure is often characterized by the mismatch between vacuum energy density in Minkowski spacetime and the observed value of dark energy density. However, the Minkowski contribution can be subtracted by normal ordering or appropriate regularization of the stress-energy tensor. This point has been discussed in detail in \cite{maggiore}, which also discusses several subtleties related to non-covariant cut-offs and quadratic divergences in FLRW spacetimes; we refer the reader to this reference as we shall be adopting a similar viewpoint as far as these issues are concerned. Our main focus here would be the terms that survive once the Minkowski contribution is subtracted. These surviving terms would involve {\it three} length scales: a UV scale $\lp$, an IR scale $\lr$, and the curvature scale $\ell_\Rcomb$, and hence one expects the resultant expression to be of the form: 
\begin{eqnarray}
\rhov &=& 
%\langle \texttt{M} |T_{ab}| \texttt{M} \rangle - \langle \texttt{M} |T_{ab}| \texttt{M} \rangle_{\bm \eta}
%\nn \\
%&=& 
\frac{\pi \hbar c}{2} 
\Biggl[ 
\frac{c_0}{\ell_\Rcomb^2 \lp^2} + \frac{c_1}{\ell_\Rcomb^2 \lr^2}  
+ \frac{c_2}{\lr^4} + \ldots
\Biggl]
\label{eq:symbolic}
\end{eqnarray}
where $c_{0,1,2}$ are $O(1)$ numerical constants which we will determine from our analysis. It is natural that 
\begin{eqnarray}
\lr \gae \ell_\Rcomb \gg \lp
\end{eqnarray}
and hence the various terms on RHS of \eq{eq:symbolic} are in descending order of magnitude. The $``\ldots"$ represents sub-dominant terms, in a sense which will become clearer in the course of the calculation.

It is our purpose in this paper to show that: (i) an expression of the form \eq{eq:symbolic} arises naturally in any curved spacetime in an arbitrary freely falling frame, and (ii) the curvature length scale $\ell_\Rcomb = 1/\sqrt{|\Rcomb|}$ is determined by a specific combination of Ricci tensor components $\Rcomb$ which depends on the specific choice of the free-fall vacuum.

%(3) many aspects of vacuum energy often studied within the framework of FLRW cosmology are in fact special cases of the above result.

We discuss certain implications and speculations in the end, in particular what the analysis does and does {\it not} say about effects often believed to be deeply connected with the structure of the quantum vacuum, such as horizon entropy and the cosmological constant (CC). These are important issues, some of which have been discussed at length in the literature (for example, see \cite{maggiore} for the case of FLRW models, relevant for the CC problem). While this work has nothing to add to that discussion, it does bring to light quite clearly a few things, which we highlight in the last section. In particular, it alerts against (mis)using the equivalence principle to make statements about gravitational effects of the vacuum energy density, since it is not the modes, but their derivatives, that determine the stress-energy tensor. 
%%%%%%%%%%%%%%%%%%%%%%%%%%%%%%%%%%%%%%%%%%%%%%%%%%%%%%%%%%%%%%%%%%%%%%%%%%%%%%%%%%%%%%%
\section{Mode solutions in freely-falling frames} \label{sec:modes} 
%%%%%%%%%%%%%%%%%%%%%%%%%%%%%%%%%%%%%%%%%%%%%%%%%%%%%%%%%%%%%%%%%%%%%%%%%%%%%%%%%%%%%%%
Since our main aim is to find the dominant correction to standard Minkowski vacuum modes in arbitrary curved spacetime, the most natural frames to choose for quantisation are the freely falling frames characterised by their geodesic tangent vector field $\bm U(\tau)$, where $\tau$ is the proper time along the trajectory. The coordinates best suited for such frames are the Fermi normal coordinates (FNC) $(\tau, x^{\mu})$, which uniquely identify any event $\mathcal P$ in a geodesically convex neighbourhood of the trajectory. These coordinates are therefore a natural analogue of the Minkowski coordinates $(T, X^{\mu})$ on which the conventional quantisation of fields in Minkowski spacetime is based. 
%%%%%%%%%%%%%%%
%%%%%%%%%%%%%%%
%The FNCs, $(\tau, x^{\mu})$, can be constructed by Fermi-Walker transport of the orthonormal tetrad along the observer worldline. 

The metric in FNCs, to $O\l[R \tau^2, R x_\mu^2, (\nabla R) x_\mu\r]$, can be written as (see, for e.g., \cite{fnc-refs})
%\begin{widetext}
\begin{eqnarray}
\DM s^2 = 
&-& \left[ \left( ~ 1 +  a_\mu x^\mu ~ \right)^2 + {\mathcal S}_{\mu \nu} x^{\mu} x^{\nu} \right] ~
\DM \tau^2 
\nn \\
&+& 2 \left[ - {\tiny{\frac{2}{3}}} \varepsilon_{\rho \alpha \nu} {\mathcal B}_{\rho \mu} x^{\mu} x^{\nu} \right] ~ \DM \tau \DM x^{\alpha} 
\nn \\
&+&
\left[ \delta_{\alpha \beta} - \frac{1}{3} \varepsilon_{\rho \alpha \mu} \varepsilon_{\sigma \beta \nu} {\mathcal E}_{\rho \sigma} x^{\mu} x^{\nu}  \right] ~
\DM x^{\alpha} \DM x^{\beta}
\label{eq:fnc-metric}
\end{eqnarray}
%\end{widetext}
where 
\begin{eqnarray}
{\mathcal S}_{\alpha \beta} &=& R_{0 \alpha 0 \beta} = {\mathcal S}_{\beta \alpha}
\\
{\mathcal E}_{\alpha \beta} &=& \frac{1}{4} \varepsilon_{\alpha \gamma \sigma} \varepsilon_{\beta \lambda \mu}
R_{\gamma \sigma \lambda \mu} = {\mathcal E}_{\beta \alpha}
\\
{\mathcal B}_{\alpha \beta} &=& \frac{1}{2} \varepsilon_{\alpha \gamma \sigma} R_{0 \beta \gamma \sigma}
\label{eq:fncdefns}
\end{eqnarray}
We also state the following useful identities, which we will repeatedly use (and refer simply as \textit{I}, when needed):

%$\left.\begin{tabular}{l}
\textit{I}: $\left\{\begin{tabular}{l}
${\mathcal S}^{\alpha}_{\phantom{\alpha} \alpha} = - R^0_{\phantom{0}0}$
\\
$R_{\rho \mu \rho \nu} = R_{\mu \nu} + S_{\mu \nu}$; \hspace{0.1cm} $R_{\rho \mu \rho \mu} = - 2 G^0_0 = 2 {\mathcal E}^\alpha_{\phantom{\alpha} \alpha}$
\\
$R = R_{\mu \mu} - {\mathcal S}_{\mu \mu}$\end{tabular}\right.$

In what follows, we will write the metric as $g^{ab}=\eta^{ab}-h^{ab}$ to keep the analysis general (and the equations transparent) for as long as possible.
%%%%%%%%%%%%%%%
%%%%%%%%%%%%%%%
%\begtn{widetext}
\begin{figure}[!htb]
%\begin{center}
\scalebox{.35}{\includegraphics{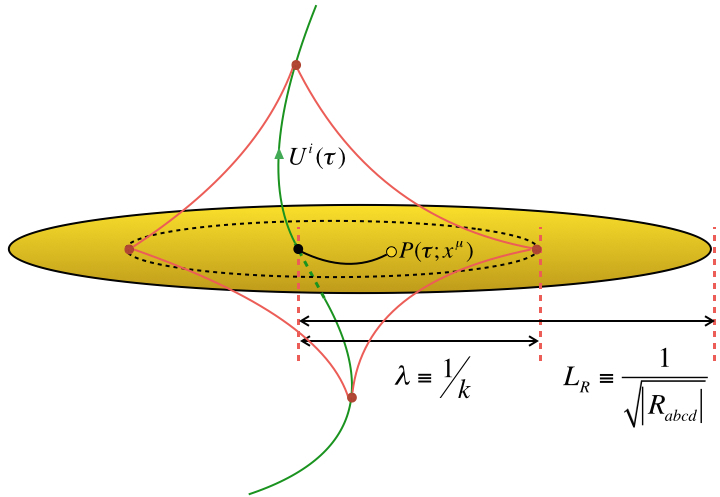}} \hfill
\caption{Relevant length scales. Red lines represent the (null) boundary of the causal diamond whose size is set by the wavelength $\lambda$ of the modes (see text for details).}
\label{fig:fnc-modes}
%\end{center}
\end{figure}
%\end{widetext}

Figure (\ref{fig:fnc-modes}) shows the set-up. Since the FNC are valid all along the trajectory, we do not have to put any restriction on the proper time intervals; the region of interest is therefore a tubular neighbourhood of the trajectory of size $\sim \lambda$. However, for future convenience, it is useful to restrict attention to the causal diamond (with it's null boundaries marked in red in the figure) based on two points on the trajectory $\Delta \tau \sim 2 \lambda$ apart. (Note that, to leading order, the corresponding forward and backward light cones would intersect in a spacelike two surface of typical size $\sim \lambda$.) 

Our aim is to solve for the field modes in FNCs, evaluate
$\langle0|T_{ab}|0\rangle U^a U^b$
in the corresponding vacuum state $|0\rangle$, and hence calculate $\rhov$ from Eq.~(\ref{eq:symbolic}). As we shall see, the dominant contribution to $\rhov$ is determined by $\Rcomb = \alpha R_{ab} U^a U^b + \beta R$, and the precise choice of $|0\rangle$ only affects the numerical constants $\alpha, \beta$.

We will now elaborate on our characterisation of the vacuum state $|\texttt{M} \rangle$. In a freely falling frame, the most natural choice of vacuum state is the one which would reduce to the Minkowski vacuum in the limit of zero curvature. Moreover, these modes must also resemble the Minkowski modes on the trajectory, since the metric is $\eta_{ab}$ all along the trajectory in FNCs. We therefore expect the relevant field modes to become the standard Minkowski plane wave modes in both these limits. With this in mind, we seek a mode solution of the Klein-Gordon equation of the form
\begin{eqnarray}
\phi_{\omega, \bm k}(x) &=& \phi^0_{\omega, \bm k}(x) \l[ 1 + f_{\omega, \bm k}(x) \r]
\nn \\
\phi^0_{\omega, \bm k}(x) &=& \frac{1}{\sqrt{k}} \exp{[i \l( \bm k \bm \cdot \bm x - \omega \tau \r)]}
\label{eq:vac-modes}
\end{eqnarray}
where $\omega=|\bm k|$ and $f_{\omega, \bm k}(x)$ will be determined perturbatively in a series expansion about the trajectory as
\begin{eqnarray}
f_{\omega, \bm k}(x) = A_{\mu} x^\mu + B_{\mu \nu} x^\mu x^\nu + C_{\mu \nu \rho} x^\mu x^\nu x^\rho + \ldots
\label{eq:Fexp}
\end{eqnarray}
The coefficients $A_\mu, B_{\mu \nu}$ etc. are functions of $\tau$, and they will also generically depend on $\omega$ and $\bm k$. These coefficients can be determined by solving the Klein-Gordon equation, $\square \phi_{\omega, \bm k}(x) =0$ order by order; some relevant details are given in the Appendix \ref{app:KGdetails}. At first order in curvature, we obtain 
\begin{eqnarray}
\l( \partial^2 + 2 {\rm i} \, k^a \partial_a \r) f_{\omega, \bm k} = h^{ab} k_a k_b - {\mathrm i} k_b \partial_a h^{ab} - {\mathrm i} k^a \partial_a h_1 
\label{eq:Feqn}
\end{eqnarray}
One can now plug the expansion (\ref{eq:Fexp}) in Eq.~(\ref{eq:Feqn}) and compare terms of same order in $x^\mu=s n^\mu$, where $s$ is the proper distance along the spacelike geodesic connecting $x^\mu$ to the trajectory and $n^\mu$ is the unit tangent to this geodesic at the point of intersection. This yields the following relations:
\begin{eqnarray}
O(s^0)&:& B_{\mu \mu} + {\mathrm i} A_{\mu} k^\mu = 0
\nn \\
O(s^1)&:& \l( -{\ddot A}_\mu + 6 C_{\rho \rho \mu} + 2 {\mathrm i} k^0 \dot A_{\mu} + 4 {\mathrm i} B_{\rho \mu} k^\rho \r) x^\mu =
\nn \\
&&\textcolor{white}{==.} i k^0 \partial_\nu h^{\nu 0} - {\mathrm i} k_\mu \partial_\nu h^{\nu \mu} - {\mathrm i} k_\mu \partial_\mu \l(\frac{h_1}{2}\r)
\nn \\
\label{eq:diffeqs}
\end{eqnarray}
and so on. Our main aim is to determine the dominant contribution of these modes to the stress-energy tensor along the trajectory. As we shall see in the next section, this contribution depends solely on $A_\mu k^\mu$, which, from the $O(s^0)$ term above, is related to the trace of $B_{\mu \nu}$. We can then use the $O(s^1)$ relation to determine a solution for $B_{\mu \nu}$, thereby completing the calculation. As we will show in the next section, the qualitative form of the resultant dominant term in $\rhov$ is universal.

%%%%%%%%%%%%%%%%%%%%%%%%%%%%%%%%%%%%%%%%%%%%%%%%%%%%%%%%%%%%%%%%%%%%%%%%%%%%%%%%%%%%%%%
\section{Stress-energy tensor} \label{sec:Tab} 
%%%%%%%%%%%%%%%%%%%%%%%%%%%%%%%%%%%%%%%%%%%%%%%%%%%%%%%%%%%%%%%%%%%%%%%%%%%%%%%%%%%%%%%
The stress-energy tensor corresponding to a scalar field $\phi(x)$ is given by 
\begin{eqnarray}
T_{ab} &=& \nabla_a \phi \nabla_b \phi \,\,\, - \frac{1}{2} g_{ab} \l( g^{cd} \nabla_c \phi \nabla_d \phi \r)
%\end{eqnarray}
\label{eq:tab} \\
%\begin{eqnarray}
T_{ab} U^a U^b &=& \l( U^a \nabla_a \phi \r)^2 + \frac{1}{2} \l( g^{cd} \nabla_c \phi \nabla_d \phi \r)
\label{eq:tuu}
\end{eqnarray}
(Note that, on the trajectory, $\bm U |_{x^\mu=0} \equiv \partial_\tau$.)
 %For a $\tau=$ constant neighbourhood of the trajectory (see Fig. (\ref{fig:fnc-modes})), we have $\bm U \equiv \l( 1/\sqrt{-g_{00}} \r) \partial_\tau$. 

We wish to calculate $\rhov=\langle \texttt{M} |T_{ab}| \texttt{M} \rangle U^a U^b$, where $|\texttt M \rangle$ is the vaccum state defined by the modes $\phi_{\omega, \bm k}(x)$ in 
Eq.~(\ref{eq:vac-modes}). Expanding the field operator in these modes in the standard manner (note that we have put in the appropriate dimensional constants for later convenience), 
\begin{eqnarray}
\hat \phi = \sqrt{\frac{\hbar c}{2}} \int \DM^3 \bm k \l( \phi_{\omega, \bm k}^* a^-_{\bm k} + \phi_{\omega, \bm k} a^+_{\bm k} \r)
\end{eqnarray}
the general expression for $\langle \texttt{M} |T_{ab}| \texttt{M} \rangle$ is given by \cite{birrell}
\begin{eqnarray}
\langle \texttt{M} |T_{ab}| \texttt{M} \rangle = \int  \DM^3 {\bm k} \, T_{ab}\l[ \phi_{\omega, \bm k}(x), \phi^*_{\omega, \bm k}(x) \r]
\end{eqnarray}
where $T_{ab}\l[\phi, \phi^* \r]$ is the expression (\ref{eq:tab}) for $T_{ab}$ treated as a bilinear form.

Although the full evaluation of RHS above is possible (albeit messy), the dominant curvature correction that survives in the limit $x^\mu \rightarrow 0$ can be easily extracted by noting that such a term can only come from the term 
\begin{eqnarray}
\delta_{\mu \nu} (\partial_\mu \phi) (\partial_\nu \phi^*)
&=& \l| \partial_\mu \phi_0 \r|^2 \l( 1 + 2 \Re[f] \r) 
%\nn \\
%&+& 
+ 2 \Re \l[ \l( \phi_0 \partial_\mu \phi_0^*\r) \partial_\mu f \r] 
\nn \\
&& + \; O(f^2, \partial f^2)
\nn 
\end{eqnarray}
since ($\Re$ denotes the Real part)
\begin{eqnarray}
\nn \\
\lim \limits_{x^\mu \rightarrow 0} \l| \partial_\mu \phi_0 \r|^2 \l( 1 + 2 \Re[f] \r) &=& k =  \lim \limits_{x^\mu \rightarrow 0} \l| \dot \phi \r|^2, \hspace{0.5cm} \mathrm{and} \hfill
\nn \\
\lim \limits_{x^\mu \rightarrow 0} \Re \l[ \l( \phi_0 \partial_\mu \phi_0^*\r) \partial_\mu f \r] 
&=& \Re \l[ |\phi_0|^2 (- i k_\mu) A_\mu \r]  
\nn \\
&=& \Re \l[ \frac{1}{k} B_{\mu \mu} \r]  
\nn \\
&=& \frac{B_{\mu \mu}}{k}
\end{eqnarray}
where we have used the first of Eqs.~(\ref{eq:diffeqs}).
Putting all of this together, we get the dominant contribution as
\begin{eqnarray}
\lim \limits_{x^\mu \rightarrow 0} \langle \texttt{M} |T_{ab}| \texttt{M} \rangle U^a U^b &=& \frac{\hbar c}{2} \int \DM^3 \bm k \, \frac{1}{k} \l( k^2 + B_{\mu \mu} \r)
\nn \\
&=& \frac{\hbar c}{2} \int \DM^3 \bm k \l( k + B_{\mu \mu}/k \r)
\nn \\
&=& (2 \pi \hbar c) \int \limits_{1/\lr}^{1/\lp} \DM k \l( k^3 + k B_{\mu \mu} \r)
\nn \\
\end{eqnarray}
The $k$ integrals are trivially done, thereby giving the final result as
\begin{eqnarray}
\rhov = \frac{\pi \hbar c}{2} 
\Biggl[ 
\frac{\epsilon}{\ell_\Rcomb^2 \lp^2} - \frac{\epsilon}{\ell_\Rcomb^2 \lr^2}  
- \frac{1}{\lr^4}
\Biggl]
\end{eqnarray}
where we have defined $1/\ell_\Rcomb^2=|2 B_{\mu \mu}|$, and $\epsilon={\rm sign}[B_{\mu \mu}]$.
which has the form mentioned in the Introduction (Eq.~(\ref{eq:symbolic})) with $c_0=\epsilon=-c_1$ and $c_2=-1$. Henceforth, we will ignore all terms except the leading first term (see comment below Eq.~(\ref{eq:symbolic})) in above expression for $\rhov$.

To complete the result, we need to solve Eqs.~(\ref{eq:diffeqs}) and find $B_{\mu \nu}$. The general structure of the solution, however, can be deduced on dimensional grounds as follows. Let us focus on the coefficients $A_\mu, B_{\mu \nu}, C_{\mu \nu \rho}$ etc. in Eq.~(\ref{eq:Fexp}). These coefficients must vanish when curvature goes to zero, since we want the modes to represent standard Minkowski vacuum everywhere in that limit. Let $\mathcal{R}$ stand for a typical curvature component. Then, on dimensional grounds, the $k$ dependence of these coefficients can be easily deduced to be
\begin{eqnarray}
\l[A_\mu\r] \equiv (\mathrm{length})^{-1} &=& \mathcal{R} k^{-1} \l( 1 + \mathcal{R}/k^2 \ldots \r)
\nn \\
\l[B_{\mu \nu} \r] \equiv (\mathrm{length})^{-2} &=& \mathcal{R} \l( 1 + \mathcal{R}/k^2 \ldots \r)
\nn \\
\l[C_{\mu \nu \rho} \r] \equiv (\mathrm{length})^{-3} &=& \mathcal{R} k \l( 1 + \mathcal{R}/k^2 \ldots \r)
\end{eqnarray}
Since we have ignored the terms quadratic in curvature in the metric, we must also do so above. Then, the leading behaviour of $B_{\mu \nu}$ can be written, quite generically, as a linear combination of $R_{\mu \nu}, R_{0 \mu 0 \nu}$, and $R \delta_{\mu \nu}$. Using the identities \textit{I}, it is easy to see that this leads to 
\begin{eqnarray}
{\epsilon}{\ell_\Rcomb^{-2}} = 2 B_{\mu \mu} = \alpha R_{ab} U^a U^b + \beta R := \mathrm{\textsf{R}}
\end{eqnarray}
which gives the form for $\rhov$ stated in the Abstract 
\begin{eqnarray}
\rhov = \l(\frac{\pi \hbar c}{2}\r) \Rcomb \, \lp^{-2}
\end{eqnarray}
As an explicit example, let us consider the following particular solution of the second equation in (\ref{eq:diffeqs}): 
\begin{eqnarray}
4 B_{\mu \nu} x^\nu &=& - \l(  \partial_\nu h^{\nu \mu} + \partial_\mu \l(\frac{h_1}{2} \r) \r)
\label{eq:condns}
\end{eqnarray}
(It is not difficult to show that this choice determines the $C_{\mu \mu \nu} k^{\nu}$ piece of the coefficient $C_{\mu \nu \rho}$.)
Using the FNC metric Eq.~(\ref{eq:fnc-metric}), %, it is then easy to show that, to $O\l[R \tau^2, R x_\mu^2, (\nabla R) x_\mu\r]$
%\begin{eqnarray}
%h^{00} &=& {\mathcal S}_{\mu \nu} x^\mu x^\nu
%\nn \\
%h^{0\mu} &=& -(2/3) R_{0 \rho \mu \sigma} x^\rho x^\sigma
%\nn \\
%h^{\mu \nu} &=& \delta_{\mu \nu} + (1/3) R_{\mu \rho \nu \sigma} x^\rho x^\sigma
%\nn \\
%\sqrt{-g} &=& 1 - (1/6) \l( R_{\mu \nu} - 2 {\mathcal S}_{\mu \nu} \r) x^\mu x^\nu
%\end{eqnarray}
this gives
\begin{eqnarray}
4 B_{\mu \nu} &=& (1/3) \l( R_{\mu \rho \nu \rho} + R_{\mu \nu} - 2 \mathcal{S}_{\mu \nu} \r)
\label{eq:Bmunutrace}
\end{eqnarray}
which (on using the identities \textit{I}) corresponds to $\alpha=1/6$ and $\beta=1/3$.
%%%%%%%%%%%%%%%%%%%%%%%%%%%%%%%%%%%%%%%%%%%%%%%%%%%%%%%%%%%%%%%%%%%%%%%%%%%%%%%%%%%%%%%
\section{Implications and Discussion} \label{sec:implications} 
%%%%%%%%%%%%%%%%%%%%%%%%%%%%%%%%%%%%%%%%%%%%%%%%%%%%%%%%%%%%%%%%%%%%%%%%%%%%%%%%%%%%%%%
What we have established above is the following: given a freely falling frame with geodesic tangent $\bm U(\tau)$, one can define a natural set of vacuum states via modes which reduce to Minkowski plane waves along the trajectory, on which the metric is $\eta_{ab}$ all along. (This is in contrast to the usual global characterisation of modes based on asymptotic structure of spacetime.) One can then calculate the leading form of the vacuum energy density after subtracting the flat space contribution, for points on/near the trajectory for which the $s \ll \lambda \ll \ell_\Rcomb$; this is given by
\begin{eqnarray}
\rhov &=& \l(\frac{\pi \hbar c}{2}\r) \Rcomb \, \lp^{-2} = \l(\frac{\pi \hbar c}{2}\r) \frac{\epsilon}{\ell_\Rcomb^2 \, \lp^{2}}
\end{eqnarray}
with $\mathrm{\textsf{R}} = \alpha R_{ab} U^a U^b + \beta R$. 

We now discuss a few implications of this result, and connect with some known results from quantum field theory in curved spacetime.

%%%%%%%%%%%%%%%%%%%
%\subsection{Cosmology}
%\textit{Cosmology}:
%%%%%%%%%%%%%%%%%%%
The above form for vacuum energy density has already been noticed in the context of cosmology \cite{geom-mean-rel, maggiore} (see also, \cite{gx}), since the observed value of dark energy density scales as 
$\rho_\textsf{DE} \sim H^2/{\ell_P}^2$, where $H$ is the Hubble length and $\ell_P=\sqrt{G\hbar/c^3}$ is the Planck length. Since curvature for FLRW models scales as $H^2$, and we expect $\lp = O(1) \ell_P$, this expression is qualitative similar to $\rhov$. Therefore, the results in cosmology seem to be just a special case of the more general result we have derived for arbitrary freely falling frames in arbitrary curved spacetimes. To further explore the implications of our result for various aspects of the cosmological constant problem, we can go a step further and investigate our expression for $\rhov$ in cases where the background curved spacetime is a solution of Einstein equations with source(s) as perfect fluid(s) having equation(s)-of-state $p_A=w_A \rho_A$ in the frame defined by $U^a$ \cite{comment1}.

As already observed before, the values of $\alpha$ and $\beta$ depend on the specific choice of solutions to Eqs.~(\ref{eq:diffeqs}) (i.e., on details of the vacuum modes). Let us first consider the special case $\beta=\alpha/4$. In this case, 
\begin{eqnarray}
\Rcomb = \alpha \left(R_{ab} - \frac{1}{4} R g_{ab}\right) U^a U^b \hspace{1cm} (\mathrm{for~} \beta=\alpha/4) 
\end{eqnarray}
which is invariant under the shift $R_{ab} \rightarrow R_{ab} + \Lambda g_{ab}$, and therefore vanishes for spacetimes in which $R_{ab}=\Lambda g_{ab}$. Therefore, for this choice of state, any bulk $\Lambda$ will not contribute to $\rhov$. More generally, it is easy to show that 
\begin{eqnarray}
\Rcomb &=& \l(\frac{8\pi G}{c^4}\r) \sum \limits_{A} \l[ \l(\frac{1+3w_A}{2}\r) \alpha + \l(1-3 w_A\r) \beta \r] \rho_A
\nn \\
&=& \l(\frac{8\pi G}{c^4}\r) \Biggl[ \l(\frac{\alpha}{2}+\beta\r) \rho_{\rm matter} + \alpha \rho_{\rm radiation} 
\nn \\
&& \textcolor{white}{\l(\frac{8\pi G}{c^4}\r) \Biggl[ \l(\frac{\alpha}{2}+\beta\r) \rho_{\rm matter}} - \l(\alpha-4\beta \r)\rho_\Lambda \Biggl]
\label{eq:eos-form}
\end{eqnarray}
which, for $\beta=(1/4)\alpha$, reduces to $\Rcomb \propto \sum \limits_A(1+w_A) \rho_A$; as mentioned above, $w_A=-1$ does not contribute in this case. In general, the values of $\alpha, \beta$, would determine the contribution of various sources such as dust ($w_A=0$) and radiation ($w_A=1/3$) to $\rhov$. This suggests that, if one were to identify the dark energy density with vacuum energy density, the contribution of various sources to it's magnitude would depend on the details of the free-fall vacuum. 
%See Appendix \ref{app:Fvacuum} for an example of a vacuum state modeled after the Bunch-Davies vacuum used in cosmology. 
As an aside, note that, from the last of Eqs.~(\ref{eq:eos-form}), it follows quite generically that $\rhov$ would change sign when (ignoring $\rho_{\rm radiation}$)
\begin{eqnarray}
\frac{\rho_\Lambda}{\rho_{\rm matter}} = \frac{\alpha/2+\beta}{\alpha-4\beta}
\end{eqnarray}
It is unclear if this says anything deeper about the {\it why-now} part of the cosmological constant problem. In particular, while speculating on any connection with the CC, one must add that what we have here is only an analysis for the scaling of the vacuum energy density, and have said nothing about {\it pressure} and hence the {\it equation of state}. As already mentioned in the Introduction (last paragraph), this work has nothing to add to discussions on such issues. 

Finally, we briefly comment on possible relevance of our result for horizon entropy. Note that it is the $\lp^{-2}$ scaling of $\rhov$ that, in the context of cosmology, allows us to connect it with the observed value of dark energy density. There is another extremely important result in quantum field theory in curved spacetime that also is often considered as having something do with the structure of quantum vacuum. This is the so called Bekenstein-Hawking entropy of a black hole of horizon radius $\mathcal A_{\rm H}$: $S(\mathcal A_{\rm H}) = \mathcal A_{\rm H}/4{\ell_{\rm p}^2} $ which scales as area rather than volume, and depends quadratically on the Planck length $\ell_{\rm p}$. It would therefore be natural to ask whether our expression for vacuum energy density $\rhov$ in freely falling frames has anything to say about such a scaling of entropy. In this context, it is interesting to mention reference \cite{jacobson-parentani}, where the authors discuss regularization of black hole entropy in (radially) freely falling frames in spherically symmetric black hole spacetimes and arrive at a non-trivial (non-quadratic) scaling with the cut-off. Essentially, the surface gravity of the Killing horizon provides an additional length scale, which makes the result in the freely fall frame non-trivial. Our analysis suggests a way of exploring such a scaling in a much more general manner, for local Rindler horizons in arbitrary curved spacetimes. This is work under progress.

%Finally, we must note that, if $\rhov$ is taken as a gravitating source (which it should), one can add it to the classical stress-energy tensor $T^{C}_{ab}$ on the RHS of Einstein equations; for example,
%\begin{eqnarray}
%G_{ab} U^a U^b = \l(\frac{8\pi G}{c^4}\r) \l( T^{C}_{ab} U^a U^b  + \rhov \r)
%\end{eqnarray}
%One can also write the remaining Einstein equations in a similar manner by calculating the pressure $p_0$ as well (along the lines discussed in \cite{maggiore} for FLRW models). Full implications of this are worth exploring further.
%%%%%%%%%%%%%%%%%%%%%%%%%%%%%%%%%%%%%%%%%%%%%%%%%%%%%%%%%%%%%%%%%%%%%%%%%%%%%%%%%%%%%%%
{\it Acknowledgements --} I thank T. Padmanabhan for reading through the manuscript, and for many helpful comments and criticisms.
%%%%%%%%%%%%%%%%%%%%%%%%%%%%%%%%%%%%%%%%%%%%%%%%%%%%%%%%%%%%%%%%%%%%%%%%%%%%%%%%%%%%%%%%%%%
%%%%%%%%%%%%%%%%%%%%%%%%%%%%%%%%%%%%%%%%%%%%%%%%%%%%%%%%%%%%%%%%%%%%%%%%%%%%%%%%%%%%%%%%%%%
\appendix 
%%%%%%%%%%%%%%%%%%%%%%%%%%%%%%%%%%%%%%%%%%%%%%%%%%%%%%%%%%%%%%%%%%%%%%%%%%%%%%%%%%%%%%%
\begin{widetext}
\section{Derivation of Eqs.~(\ref{eq:Feqn}) and (\ref{eq:diffeqs})} \label{app:KGdetails}

Using the FNC metric Eq.~(\ref{eq:fnc-metric}), it is easy to read off the following coefficients, to $O\l[R \tau^2, R x_\mu^2, (\nabla R) x_\mu\r]$
\begin{eqnarray}
h^{00} &=& {\mathcal S}_{\mu \nu} x^\mu x^\nu
\nn \\
h^{0\mu} &=& -(2/3) R_{0 \rho \mu \sigma} x^\rho x^\sigma
\nn \\
h^{\mu \nu} &=& \delta_{\mu \nu} + (1/3) R_{\mu \rho \nu \sigma} x^\rho x^\sigma
\nn \\
\sqrt{-g} &=& 1 - (1/6) \l( R_{\mu \nu} - 2 {\mathcal S}_{\mu \nu} \r) x^\mu x^\nu
\nn \\
&=& 1 + h_1
\label{eq:fnc-metric-appendix}
\end{eqnarray}

Using above, and our ansatz (\ref{eq:vac-modes}) for the modes, we can show that the Klein-Gordon equation $\square \phi=0$, to relevant order, reduces to
%\begin{widetext}
\begin{eqnarray}
\square \phi &=& \l(1+f\r) \l(\omega^2 - \bm k^2\r) + 2 i k^a \partial_a f + \partial^2 f - h^{ab} k_a k_b + i k_b \partial_a h^{ab} + i k^a \partial_a h_1 = 0
\end{eqnarray}
%\end{widetext}
where $k^a=\l[\omega, \bm k \r]$ and $\partial^2 \equiv \eta^{ab} \partial_a \partial_b$. The zeroth order term (in curvature) gives $\omega^2=|\bm k|^2$, which, when plugged back, gives the first order equation Eq.~(\ref{eq:Feqn}). One can then plug Eqs.~(\ref{eq:fnc-metric-appendix}) in Eq.~(\ref{eq:Feqn}), and compare terms of same order in $x^\mu=s n^\mu$ to obtain Eqs.~(\ref{eq:diffeqs}).
\end{widetext}


\begin{thebibliography}{99}
\bibitem{comment1}
The generalisation to the case where different fluids have different relative velocities with respect to $U^a$ can be handled in a straightforward manner. However, it can introduce new features which are worth investigating further.
\end{thebibliography}
\end{document}